\documentclass[a4paper,fleqn]{cas-dc}

\usepackage[numbers]{natbib}
\usepackage{braket}
\usepackage{amsmath}
\usepackage{multirow}
\usepackage{array}
\usepackage{siunitx}

\def\tsc#1{\csdef{#1}{\textsc{\lowercase{#1}}\xspace}}
\tsc{WGM}
\tsc{QE}
\tsc{EP}
\tsc{PMS}
\tsc{BEC}
\tsc{DE}

\begin{document}
\let\WriteBookmarks\relax
\def\floatpagepagefraction{1}
\def\textpagefraction{.001}
\shorttitle{}
\shortauthors{M. Shaban et~al.}

\title [mode = title]{Towards Quantum Network Performance Metrics: Challenges and Demonstration}                      




\author[1,2]{Mohamed Shaban}
\cormark[1]
\ead{mshaban@tntech.edu}


\author[3]{Mariam Kiran}
\ead{kiranm@ornl.gov}

\author[1]{Muhammad Ismail}
\ead{mismail@tntech.edu}

\affiliation[1]{organization={Cybersecurity Education, Research, and Outreach Center (CEROC), Tennessee Tech University},
                city={Cookeville},
                postcode={38505}, 
                state={Tennessee},
                country={USA}}

\affiliation[2]{organization={Department of Mathematics, Faculty of Education, Alexandria University},
                city={Alexandria},
                country={Egypt}}

\affiliation[3]{organization={Quantum Communications and Networking, Oak Ridge National Laboratory},
                city={Oak Ridge},
                postcode={37830}, 
                state={Tennessee},
                country={USA}}

\cortext[cor1]{Corresponding author}



\begin{abstract}
As quantum networks move toward practical deployment, standardized performance monitoring becomes essential. This article proposes a structured monitoring framework for quantum networks with performance metrics, including quality (e.g., entanglement fidelity, QBER, loss, dark count rate), throughput and latency (e.g., entanglement rate, waiting time), timing (e.g., coincidence window, production and coincidence jitter), and exogenous factors (e.g., temperature, humidity, vibrations). These measurements enable real-time observability, benchmarking, and control, supporting use cases such as fault diagnosis, adaptive timing, and entanglement routing. Additionally, we implement a non-invasive prototype environmental monitoring system integrated with the quantum network infrastructure at Oak Ridge National Laboratory, demonstrating practical feasibility of live data collection and alert generation. Furthermore, we discuss the challenges of real-time monitoring and the trade-offs between observability and system performance. This work establishes a foundation for developing advanced quantum network monitoring systems and lays the groundwork for future autonomous control and quantum software-defined networking.
\end{abstract}





\maketitle

\section{Introduction}
Quantum networks are essential to enable critical applications, such as distributed quantum computing, distributed quantum sensing, secure communication, and quantum key distribution (QKD) \cite{Scarani_2009,Fitzsimons2017,shettell,146}, unlike classical networks, quantum networks rely on entangled pairs to perform tasks \cite{146}, making these networks highly sensitive to environmental noise and fluctuations, making performance monitoring challenging.

Here, we present a comparison of traditional classical and quantum performance metrics, such as latency, packet loss, and throughput and how to capture the behavior of quantum communications. Further, for example, entanglement fidelity quantifies how closely a shared quantum state approximates the ideal maximally entangled state, does not exist in classical networks. Classical communication links carry bits, not entangled quantum states, so classical network quality is measured by metrics like throughput (bits/sec), latency, or packet loss, none of which capture the quantum correlation quality that fidelity measures. 

Classical and quantum networks even define throughput differently. In classical networks, throughput reflects the raw data transmission rate. In contrast, quantum networks define throughput as the number of entangled pairs delivered per second at or above a specified fidelity threshold \cite{10882978}. In other words, a quantum link is not characterized only by the number of qubits transmitted, but by the quality of the entanglement those qubits maintain. Also, the latency in quantum networks cannot be defined as the time it takes for data to travel across the network, as described in classical networks. This is because generating an entangled pair is a probabilistic process that requires several trials and depends on whether entanglement purification (also known as distillation) is achieved, and on how many pairs are used during purification. The waiting time for the communicating parties to receive an entangled pair is a random variable and cannot be characterized by a fixed value. In contrast, latency in classical networks is generally treated as a nearly deterministic quantity. It is typically broken down into four components: propagation delay (the time for signals to physically travel through the medium), transmission delay (the time to push bits onto the communication link), processing delay (at intermediate routers or switches), and queuing delay (due to network congestion). Among these, propagation and transmission delays are essentially fixed for a given path and packet size, while queuing delay introduces variability. Additionally, with quality-of-service (QoS) mechanisms, this variability can be minimized or bounded, making overall classical network latency predictable in practice \cite{1354590}. Thus we define key performance indicators in quantum networks and effective means to monitor them in real-time, needed to enable an intelligent quantum network management. 

\subsection{Related Work}
\begin{figure}
    \centering
    \includegraphics[width=0.9\linewidth]{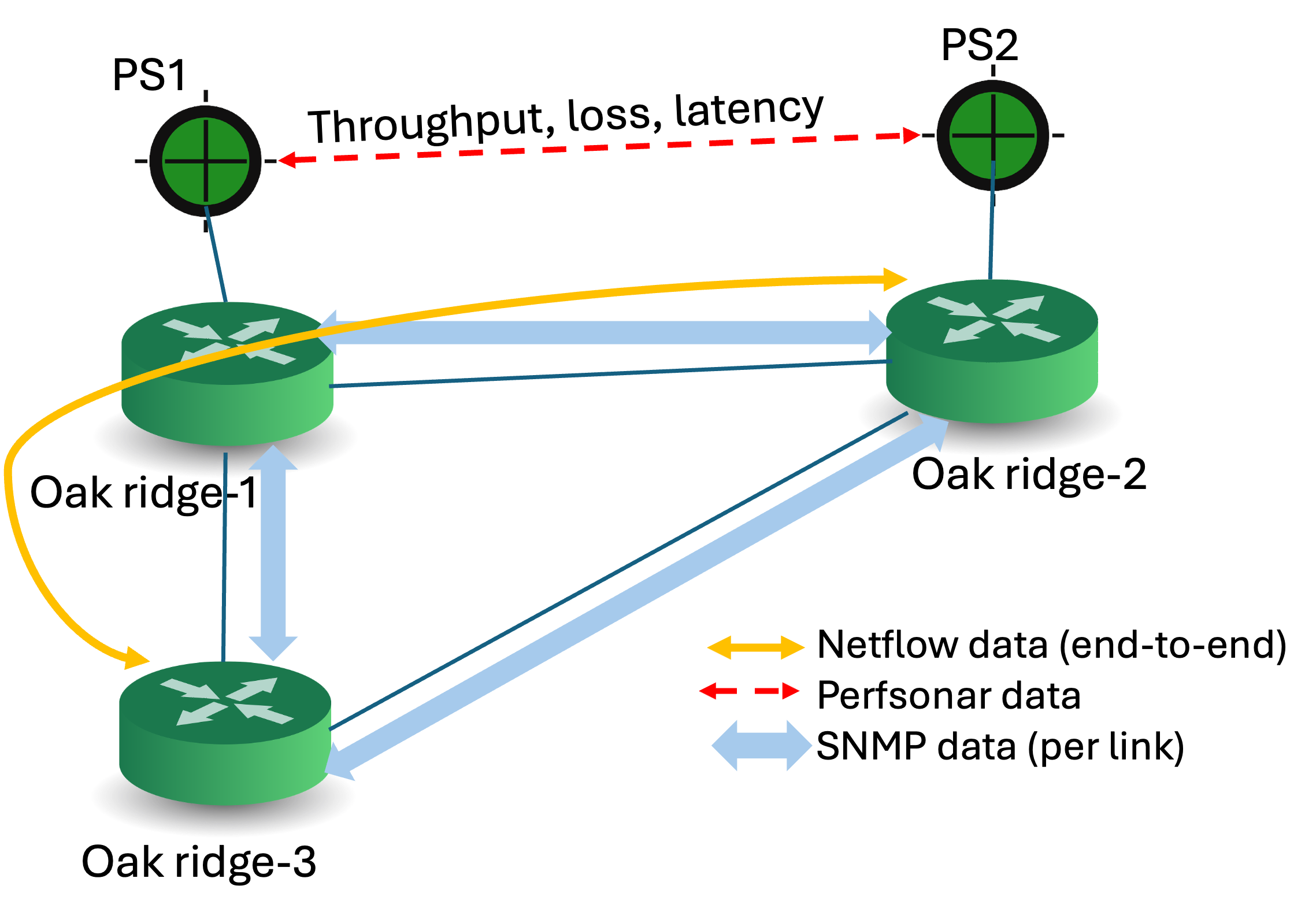}
    \caption{Monitoring in a classical network, showing a three-node topology instrumented with perfSONAR hosts (PS1 and PS2) to collect active measurement data, where NetFlow and SNMP denote flow-level traffic records exported by network devices and device/interface statistics, respectively.}
    \label{fig:cm}
\end{figure}
Network performance monitoring comprises of visualizing, monitoring, optimizing, troubleshooting and reporting on the service quality of your network. Different monitoring tools collect various data such as packet loss, flow data, throughput, and more, providing a complete picture of the network infrastructure. Figure \ref{fig:cm} shows the commonly collected network performance metrics in classical networks. Netflow data gives src id, dest id, bytes transferred and protocol. SNMP gives packet counters per router and tools like perfSonar allows one to measure throughout, latency and loss used to measure performance.

Quantum networking research has primarily focused on protocol design, such as entanglement distribution and routing algorithms, often evaluating performance through a limited set of physical-layer metrics. Commonly used metrics include entanglement fidelity, quantum bit error rate (QBER), and entanglement rate, which are typically employed to assess the correctness or efficiency of specific protocols. For instance, the work in \cite{10684482} utilized entanglement fidelity to evaluate the effectiveness of a reinforcement learning-based routing strategy. The studies in \cite{10684482} and \cite{10820730} used entanglement fidelity as a key metric to compare the performance of different quantum network architectures. The work in \cite{8068178} uses fidelity and hop count as performance indicators to identify optimal routing paths, while \cite{10.1145/3341302.3342070} presents a link-layer protocol for quantum networks and evaluates throughput and fidelity. The work in \cite{JRC132426} uses QBER to assess the feasibility and security of space-based QKD links. Additionally, the utility function introduced in \cite{10313675} incorporates both fidelity and generation rate to optimize entanglement quality and throughput. These works focus primarily on protocol evaluation and lack support for comprehensive and system-wide monitoring frameworks. 

Classical frameworks such as perfSONAR \cite{tierney2009perfsonar} rely on tools like iPerf \cite{iperf} and nuttcp \cite{nuttcp} to enable real-time observability of classical metrics like throughput, jitter, and packet loss. Quantum networks operate under different physical principles, such as the no-cloning theorem and the probabilistic nature of entanglement generation, which limit the applicability of classical monitoring techniques. These distinctions create a new set of metrics that reflect quantum-specific phenomena such as entanglement fidelity, photon generation success probability, and temporal correlations, not similar to packet delivery rates or bandwidth utilization.


\subsection{Contributions}
This paper makes the following contributions to the field of quantum networking:
\begin{itemize}
    \item We propose a standardized framework that define a comprehensive set of quantum network performance metrics across three primary categories: quality metrics, throughput and latency metrics, and timing metrics. The framework is complemented by exogenous factors, representing external environmental and operational conditions that affect network stability and performance. Together, these components provide a structured foundation for observability in quantum networks, capturing both characteristics and network-level behaviors. The framework is designed to support rigorous experimental validation as well as simulation-based evaluation.

    \item We identify the practical applications of these metrics in real-time monitoring, adaptive control, fault diagnosis, resource scheduling, benchmarking, and the development of intelligent quantum network control systems.

    \item We identify key challenges in implementing the proposed metrics, including overhead from measurements, inter-dependencies among metrics, and the complexity of control mechanisms.

    \item We implement a non-invasive prototype environmental monitoring system integrated with the quantum network infrastructure at Oak Ridge National Laboratory (ORNL). The prototype employs a Raspberry Pi, SHT35 environmental sensor, Prometheus, and Grafana to enable real-time data acquisition, visualization, and alerting without disrupting ongoing entanglement distribution experiments. This implementation demonstrates how exogenous factors can be systematically collected and incorporated into a broader quantum network observability framework.
\end{itemize}

The remainder of this article is organized as follows. Section \ref{sec:background} provides an overview of quantum networking. Section \ref{sec:metrics} introduces the key quantum network performance metrics. Section \ref{sec:applications} discusses the practical applications of these metrics. Section \ref{sec:challenges} presents the challenges involved in their implementation. Section \ref{sec:prototype} presents the implementation of an environmental monitoring prototype deployed at ORNL. Finally, Section \ref{sec:conclusion} concludes the article and outlines directions for future work.

\section{Quantum Networking Overview}
\label{sec:background}
Quantum networks enable the transmission of quantum information between distributed nodes by leveraging the fundamental properties of quantum mechanics. Unlike classical networks, where connectivity is established through physical or logical links alone, quantum network protocols require the prior distribution of shared entanglement between nodes. In this sense, the effective connection in a quantum network is established through entangled quantum states rather than solely through communication channels. Entanglement can be generated using several physical encodings, including polarization, time-bin encoding, spatial-mode (dual-rail), and frequency encoding \cite{PRXQuantum.5.010202}. Polarization entanglement exploits correlations between the polarization states of paired photons, while time-bin encoding represents quantum information in relative photon arrival times within well-defined temporal intervals. Spatial-mode and frequency encodings provide additional degrees of freedom that can improve robustness, multiplexing capability, and compatibility with different physical platforms. Regardless of the encoding method, these approaches generate entangled photon pairs that serve as the fundamental resource for executing quantum networking protocols such as quantum key distribution, teleportation, and distributed quantum computation. A typical quantum network consists of multiple nodes, each equipped with several core components. The primary elements include entangled photon sources and single-photon detectors. The photon source generates entangled pairs at a defined production rate, typically measured in entangled pairs per second. Ideally, both photons from a pair are detected simultaneously at their respective nodes, preserving temporal correlation. In practice, however, unequal fiber lengths, dispersion, detector jitter, and hardware imperfections introduce arrival-time mismatches. To address these timing uncertainties, a coincidence window is defined, representing a narrow time interval within which detection events are considered correlated. Accurate coincidence detection relies on precise time synchronization across all network nodes. As quantum networks scale beyond point-to-point links, additional components become necessary. Quantum memories provide short-term storage of quantum states, enabling coordination across asynchronous events and supporting entanglement swapping and purification protocols. Quantum repeaters integrate memory, entanglement generation, and measurement capabilities to extend communication distances beyond the limits imposed by optical attenuation and loss. Together, these components form the architectural foundation for scalable, multi-hop quantum networks.

\begin{figure*}
    \centering
    \includegraphics[width=1\linewidth]{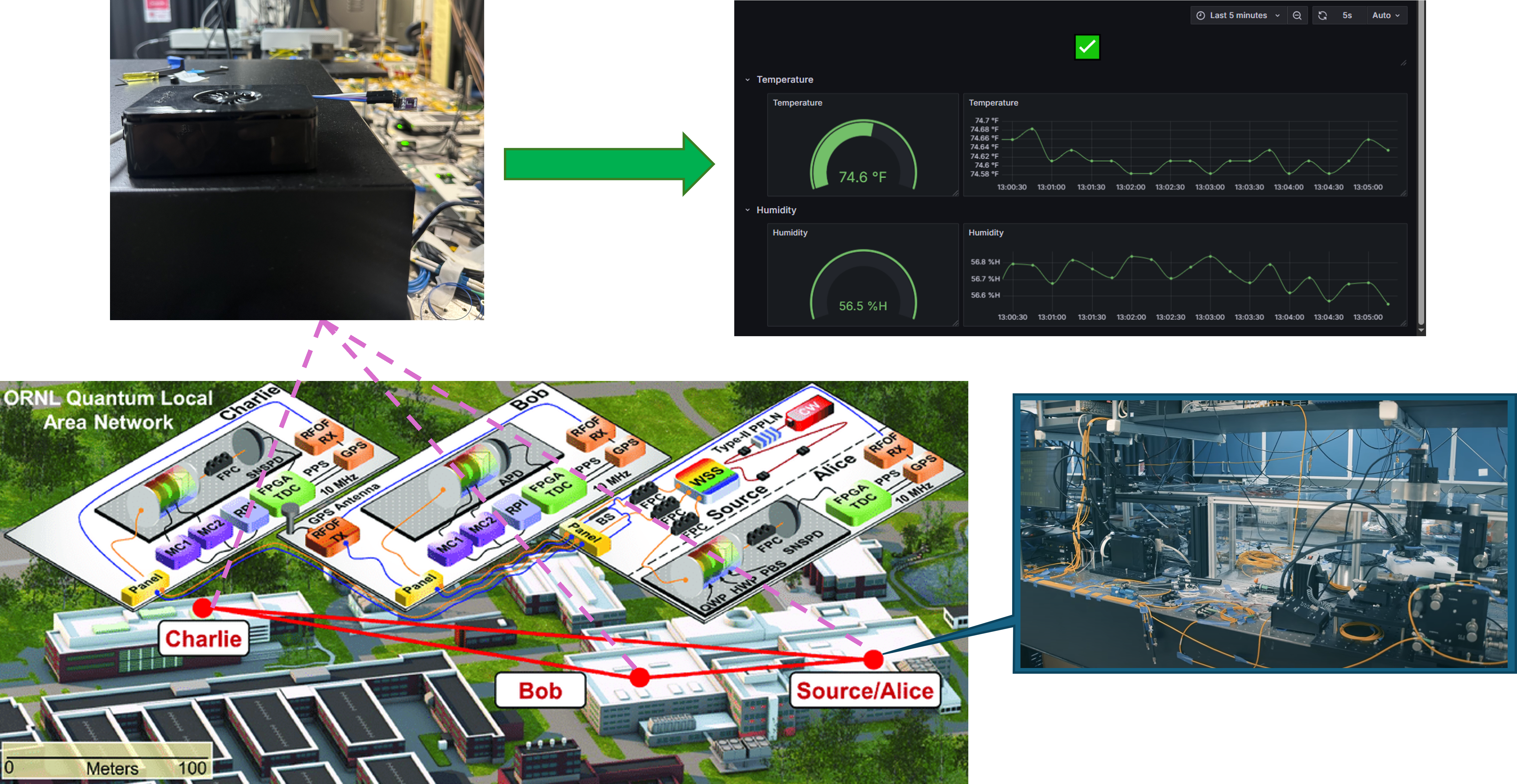}
    \caption{Illustration of the ORNL quantum local area network, spanning three interconnected quantum nodes labeled Alice, Bob, and Charlie. The central diagram illustrates the architectural layout and major components of each node. A zoomed-in laboratory image highlights the physical quantum optics setup at one site. Each laboratory includes a prototype monitoring system based on a Raspberry Pi 5 connected to an SHT35 digital sensor. These systems currently collect real-time temperature and humidity data, which are pushed to a Prometheus database and visualized on a Grafana dashboard. The illustration in the lower-left panel is adapted from \cite{PRXQuantum.2.040304}, licensed under the Creative Commons Attribution 4.0 International (CC BY 4.0) License (\url{https://creativecommons.org/licenses/by/4.0/}).}
    \label{fig:network}
\end{figure*}

\noindent\textbf{ORNL Quantum Network.} At ORNL, we have access to a quantum networking infrastructure that supports real-world experimentation and testbed validation. The network integrates several critical components to enable entanglement-based quantum communication. The process begins with a laser source that generates photons, which are directed into a periodically poled lithium niobate (PPLN) crystal. The PPLN performs type-II spontaneous parametric downconversion, producing spectrally correlated and polarization-entangled photon pairs in the ideal Bell state. These entangled photons are routed through a wavelength-selective switch (WSS), which divides the entangled qubits among the network parties. Each WSS output is connected to fiber polarization controllers for polarization alignment before reaching the polarization analysis modules at the receiver nodes. These modules include half-wave plates (HWPs), quarter-wave plates (QWPs), and polarizing beamsplitters (PBSs) for polarization state analysis. The HWPs and QWPs are programmatically adjusted using motion controllers (MCs) and Raspberry Pi for precise manipulation. The photons are detected using avalanche photodiodes (APDs) or superconducting nanowire single-photon detectors (SNSPDs), with detection events timestamped by field-programmable gate array (FPGA)-based time-to-digital converters (TDCs). Timing synchronization across the nodes is maintained with white rabbit nodes (WRNs) and a white rabbit switch (WRS), ensuring sub-nanosecond clock alignment. An arbitrary waveform generator (AWG) aids in generating synchronized control signals. Finally, the collected data is processed and analyzed in real time using a coincidence counting (Coinc) system. A picture of the quantum networking setup at ORNL is shown in Figure \ref{fig:network}.

\section{Quantum Network Performance Metrics}
\label{sec:metrics}
\vspace{0.5mm}
Quantum entanglement serves as the actual connection between nodes in quantum networks and enables most quantum network protocols \cite{10684482,PhysRevLett.70.1895,8910635,146}. Accordingly, some of the performance metrics presented in this section, such as entanglement fidelity and entanglement rate, are naturally defined in terms of entangled states. However, these metrics can be generalized to broader quantum services. For example, fidelity can be used to assess the quality of any quantum state, not just entangled ones, and the rate metric can be adapted to quantify protocol throughput, such as the secret key rate in QKD. This flexibility allows the proposed framework to support a wide range of quantum applications beyond entanglement distribution. The performance metrics defined across this section provide a structured foundation for evaluating, monitoring, and optimizing quantum network behavior under real-world conditions. Table \ref{tab:metrics} presents an overview of these metrics along with brief definitions.

\begin{table*}[]
    \renewcommand{\arraystretch}{1.3}
    \centering
    \caption{Summary of Quantum Network Performance Metrics.}
    \begin{tabular}{|m{2.5cm}|m{3.4cm}|m{9.5cm}|}
    \hline
    \textbf{Category} & \textbf{Metric} & \textbf{Description}\\\hline
    \multirow{4}{*}{\raisebox{0mm}{Quality Metrics}} 
    & Entanglement Fidelity & The closeness of a shared quantum state to an ideal entangled state.\\
    \cline{2-3}
    & QBER &Fraction of qubits with incorrect measurement outcomes.\\
    \cline{2-3}
    & Loss & Proportion of quantum signals lost during transmission.\\
    \cline{2-3}
    & Dark Count Rate & False photon detections in the absence of actual photons.\\\hline

    \multirow{2}{*}{\raisebox{-8mm}{\shortstack[c]{Throughput and\\ Latency Metrics}}}
    &Entanglement Rate & The number of successfully distributed entangled pairs per second.\\
    \cline{2-3}
    &Photon Count Rate & The number of photon detection events per second at a detector.\\
    \cline{2-3}
    &Waiting Time & Time between request initiation and service fulfillment.\\
    \hline

    \multirow{3}{*}{\raisebox{0mm}{Timing Metrics}}
    & Coincidence Window & Time interval for identifying correlated photon detections. \\
    \cline{2-3}
    & Production Jitter & Variability in time intervals between entangled pair generations. \\
    \cline{2-3}
    & Coincidence Jitter & Timing variation between detections of photons from the same pair.\\
    \hline

    \multirow{4}{*}{\raisebox{0mm}{\shortstack[c]{Exogenous Factors}}}
    & Source Temperature & Temperature of the entangled photon source, affecting stability. \\
    \cline{2-3}
    & Room Temperature & Ambient environmental temperature affecting device performance. \\
    \cline{2-3}
    & Moisture & Humidity level affecting communication and optical components. \\
    \cline{2-3}
    & Vibrations & Mechanical disturbances affecting alignment and signal stability. \\
    \hline

    \hline
    \end{tabular}
    \label{tab:metrics}
\end{table*}

\subsection{Quality Metrics}
This section introduces metrics that evaluate the correctness, integrity, and noise level of distributed quantum states, such as entanglement fidelity, QBER, loss, and dark count rate.

\subsubsection{Entanglement Fidelity:}
Fidelity measures the similarity between two quantum states. Entanglement fidelity is a specific form of fidelity that quantifies the similarity between a distributed entangled state and the ideal maximally entangled state. It reflects the quality of distributed entanglement and directly impacts the reliability of quantum network protocols. In quantum networks, entangled pairs serve as the primary resource enabling a range of protocols, including quantum teleportation, superdense coding, QKD, and distributed quantum computing and sensing. The fidelity of these pairs directly impacts the success probability, security, and computational correctness of such operations. Formally, the fidelity $F$ is defined as \cite{136}
\begin{equation}
    \label{eq:fidelity}
    F = \bra{\psi} \rho \ket{\psi},
\end{equation}
where $\rho$ is the density matrix of the distributed state and $\ket{\psi} = \ket{\Phi^{+}}$ represents the ideal target state. In the case of entanglement fidelity, $\ket{\psi}$ is typically taken to be the maximally entangled Bell state ($\ket{00}+\ket{11})/\sqrt{2}$. The entanglement fidelity ranges from $0$ to $1$, where $F=1$ indicates perfect agreement with the ideal state, and lower values reflect degradation due to noise, loss, or operational imperfections. In practical quantum systems, maintaining high fidelity is essential for reliable and secure communication. For example, in QKD protocols, fidelity below a certain threshold may compromise security or cause key mismatches between communicating parties. In quantum identity authentication protocols, reduced fidelity can lead to authentication failures, where malicious activities may go undetected, or legitimate sessions may be falsely flagged as attacks. Moreover, entanglement fidelity not only serves as a quality assurance metric for entangled link setup but also informs adaptive strategies such as link reconfiguration, entanglement purification, or routing decisions in multi-hop quantum networks. In experimental platforms, fidelity is typically estimated via quantum state tomography or statistical Bell tests performed over multiple distributed entangled pairs. In simulations, fidelity is computed using the density matrix of the quantum state under modeled noise conditions.

\subsubsection{Quantum Bit Error Rate (QBER)}
It is a critical metric in quantum communication systems that quantifies the fraction of qubits whose measured values differ from their expected values. It serves as a direct indicator of the noise and error present in a quantum channel. Formally, QBER is defined as
\begin{equation}
    QBER= \frac{n_{e}}{n_{t}},
\end{equation}
where $n_e$ is the number of incorrect measurement outcomes, and $n_t$ is the total number of qubits compared. Several factors can contribute to the QBER, including channel noise, polarization drift, optical misalignment, detector dark counts, and potential eavesdropping attempts. In the context of QKD, QBER is often used to assess the security of the key generation process and to determine whether eavesdropping or hardware imperfections are corrupting the transmitted quantum information. The QBER is typically monitored continuously during the key exchange process. If it exceeds a predefined security threshold, the protocol is aborted, as this may indicate the presence of an eavesdropper or significant channel noise. Beyond QKD, QBER is also relevant in distributed quantum computing and entanglement-based communication, where high QBER can corrupt non-local quantum gates or lead to incorrect teleportation outcomes. Lower QBER values indicate higher integrity and can allow for more efficient error correction and privacy amplification. Lower QBER values indicate higher integrity and can allow for more efficient error correction and privacy amplification. Monitoring QBER in real time supports the dynamic management of quantum networks. It enables quality control of links, triggers purification procedures when necessary, and serves as a feedback signal for entanglement routing and resource allocation policies. In practice, QBER is estimated by sending known qubit states and comparing measured results to expected outcomes. In simulations, QBER is computed by tracking the number of erroneous measurement outcomes induced by simulated noise and imperfections.

\subsubsection{Loss} 
It refers to the proportion of quantum signals, typically photons, that fail to reach their intended destination during transmission through a quantum channel. It is one of the most critical sources of performance degradation because it directly impacts other key metrics, including entanglement fidelity, QBER, entanglement rate, and waiting time. As such, loss affects not only link-level performance but also the overall reliability of quantum protocols such as entanglement distribution, quantum teleportation, and QKD. For instance, the presence of high loss reduces the probability of successfully receiving entangled pairs, thereby lowering the entanglement rate. As more attempts are required to achieve a successful transmission, the waiting time for quantum services increases. Simultaneously, low signal strength increases the relative impact of dark counts, contributing to higher QBER. It is important to distinguish between two related quantities: channel attenuation and system efficiency. Channel attenuation represents the optical loss introduced by the transmission medium itself, such as fiber attenuation, atmospheric absorption, or beam divergence, and is typically expressed in decibels (dB) as
\begin{equation}
    \text{Channel Attenuation} = -10 \log_{10}{\frac{p_{out}}{p_{in}}},
\end{equation}
where $p_{in}$ and $p_{out}$ are the optical powers at the input and output of the channel, respectively. In contrast, system efficiency accounts for additional factors beyond the channel, including detector quantum efficiency, optical coupling efficiency, filter insertion loss, and polarization-control losses. Distinguishing between these two quantities allows for more accurate diagnosis of performance degradation, isolating whether losses arise from the transmission path or the system hardware. Consequently, loss can be expressed probabilistically as
\begin{equation}
    \text{Loss Probability} = 1 - \eta,
\end{equation}
where $\eta$ denotes the transmissivity of the quantum channel, defined as the ratio of photons successfully detected at the receiver to the total number of photons emitted by the source. Common sources of loss include attenuation, scattering, and bending in fiber-based systems, as well as beam divergence, turbulence, and atmospheric attenuation in free-space optical (FSO) channels. Additionally, alignment errors and imperfections in optical components introduce inherent transmission losses. Monitoring loss facilitates network debugging by identifying issues such as misalignments, faulty components, or degraded links, thereby enabling timely corrective actions to sustain network performance. In physical quantum systems, loss can be measured using photon counters and known source emission rates, or by continuously monitoring the optical power at the receiver. In simulation environments, loss is typically modeled using parameters derived from experimental data, including physical distance, transmission medium, and the attenuation coefficient, allowing a realistic representation of photon propagation through various quantum channels.

\subsubsection{Dark Count Rate}
It refers to the number of false detection events registered by a photon detector in the absence of actual photon arrivals. These false-positive counts are typically caused by several environmental factors, such as temperature fluctuations, background light, and cosmic radiation interacting with the detector. Dark counts are a critical source of error in quantum communication systems, especially when operating at the single-photon level, where each detection event is assumed to correspond to a meaningful quantum signal. For instance, in QKD, each detected photon contributes to the generation of secret key bits. If a dark count is mistakenly interpreted as a valid detection, it introduces a bit error, resulting in corruption of the secret key and potentially compromising the security of the protocol. Similarly, in entanglement distribution, dark counts can trigger false coincidence events, causing the system to incorrectly identify the presence of entangled pairs. This can break the correlation between genuine entangled pairs, thereby reducing the success probability of entanglement distribution. Moreover, it can mislead the fidelity measurements, as the recorded detection events may not correspond to actual entangled states. Monitoring dark count rates is essential for calibrating detection settings, such as adjusting the coincidence window to reduce false positives. Additionally, dark count monitoring can serve as an indicator of detector health, as an abnormal rise in the dark count rate may suggest hardware malfunction. In experimental settings, dark counts are measured by observing the detector output over time in the absence of an input signal. In simulations, dark count events are typically modeled as Poisson-distributed background noise added to the detection process \cite{Menkart:22}.

\subsection{Throughput and Latency Metrics}
This section presents metrics that characterize the delivery rate of quantum states (e.g. entangled pairs) and the responsiveness of the network.
\subsubsection{Entanglement Rate}
It measures the number of entangled pairs successfully distributed between two nodes per unit time. It serves as a key performance indicator for the throughput of a quantum link or path and is essential for evaluating the effectiveness and scalability of quantum networks. Formally, the entanglement rate $R_e$ can be expressed as

\begin{equation}
    R_e=\frac{n_s}{t} \;\;\;\;\;\; (e_{bits}/sec),
\end{equation}
where $n_s$ is the number of entangled pairs successfully generated between two nodes during a measurement period $t$ (in seconds). In practical systems, the entanglement rate can be controlled directly by adjusting the source pulsing frequency or modifying the width of the coincidence detection window, which influences how many entangled events are identified per unit time. However, modern quantum devices impose upper bounds on the achievable rate due to limitations in source brightness, detector recovery time, link efficiency, and channel loss. The rate may also be indirectly regulated through higher-level mechanisms such as entanglement purification protocols, which reduce the final rate while improving fidelity, or adaptive routing strategies that favor more reliable paths to maximize usable entanglement pairs. While increasing the generation rate of entangled pairs can improve throughput, it often comes at the cost of fidelity. The entanglement rate must be balanced with entanglement fidelity. Monitoring the entanglement rate is vital for benchmarking the capacity of quantum links or routing paths, allocating quantum resources efficiently, and enabling effective network control. It also supports protocol-level optimization, such as dynamically adjusting purification thresholds or re-routing traffic in response to throughput constraints. In physical systems, it is measured using timestamped coincidence detection events over a defined period. In simulation, the entanglement rate is typically computed by counting the number of successfully generated entangled pairs over a simulated time interval.

\subsubsection{Photon Count Rate}
Photon count rate represents the number of photon detection events recorded per second at a given detection channel. It provides a direct measurement of the photon flux arriving at a detector and is commonly used to characterize source brightness, optical alignment, channel loss, and detector response. In entanglement-based quantum networks, the count rates observed in the two entanglement arms influence the probability of detecting correlated photon pairs and therefore affect the achievable entanglement rate. Monitoring photon count rates helps diagnose system performance, detect alignment drift, and identify variations in experimental conditions that may impact other performance metrics such as entanglement rate and fidelity. In experimental platforms, count rates are measured directly from detector outputs and are typically expressed in Hz (photons/sec). In simulation environments, photon count rates are derived from modeled parameters such as source emission rate, channel transmissivity, detector efficiency, and background noise.

\subsubsection{Waiting Time}
Waiting time is a performance metric that captures the total duration between the initiation of a quantum service request and its successful completion. It is particularly relevant in quantum networks where key services, such as entanglement distribution, quantum teleportation, and QKD establishment, depend on probabilistic processes. Unlike nearly deterministic classical systems, quantum services often rely on entanglement generation, which may require multiple attempts before success due to channel losses, environmental noise, and source imperfections. As a result, waiting time is best modeled as a random variable whose probability distribution depends on several factors such as the characteristics of the quantum link, entanglement source, and network traffic conditions. Let $W$ denote the random variable representing the waiting time for a given request. The probability distribution $P(W \leq t)$ describes the likelihood that a request is fulfilled within time $t$. From this, nodes can estimate the expected waiting time required to receive a specific service. Waiting time is not only a metric that reflects network performance, it also serves as a decision-making tool for higher-level protocols. For instance, it can be used to design adaptive scheduling policies, where services are prioritized based on specific criteria such as urgency, request type, or resource availability. Moreover, waiting time can be used to detect congestion or bottlenecks, especially when the observed delay for certain services consistently exceeds the expected threshold. In such scenarios, the system can trigger corrective actions, such as traffic rerouting or entanglement resource reallocation. In physical testbeds, timestamps can be recorded at request initiation and service completion to estimate waiting times, enabling performance evaluation and dynamic control of the quantum network. In simulation environments, waiting times can be logged per request and analyzed to construct the probability distributions.

\subsection{Timing Metrics}
iming metrics characterize the temporal precision of entanglement generation and detection, which is essential for correctly identifying entangled events and maintaining synchronization across distributed nodes. They are typically evaluated using the full width at half maximum (FWHM) or standard deviation ($\sigma$) of time-difference histograms, with measurement cadences of $1-10$ s to balance resolution and overhead.

\subsubsection{Coincidence Window}
It is a time interval within which two photon detection events at separate detectors are considered to correspond to the same entangled pair. It serves as a temporal correlation filter used in post-processing to identify valid entanglement-generation events. Let $\Delta t_c$ denote the coincidence window duration. Detection events at detectors $A$ and $B$ occurring within $|t_A-t_B| \leq \Delta t_c$ are treated as coincident. The choice of the coincidence window $\Delta t_c$ is critical to ensuring accurate identification of entangled events. If $\Delta t_c$ is too narrow, genuine entangled pairs may be missed due to timing variations or network delays. Conversely, if it is too wide, the likelihood of accidental coincidences, such as those caused by dark counts or uncorrelated photons, increases. Since $\Delta t_c$ depends on hardware characteristics and is influenced by temporal variations such as production jitter and network delays, different links in a heterogeneous network may require different window settings. Consequently, careful calibration and real-time monitoring are important, as $\Delta t_c$ may need to be dynamically adjusted to maintain reliable identification of entanglement-generation events. In experiments, the coincidence window is calibrated using detection hardware timing offsets and adjusted empirically to optimize the balance between detection sensitivity and false coincidences. In simulations, this window can be defined as a fixed time interval within which photon detection events are matched for entanglement verification. 

\subsubsection{Production Jitter}
It refers to the variation in the time intervals between consecutive generations of entangled photon pairs by an entanglement source. Ideally, the source emits pairs at a consistent rate, supporting predictable timing and coordination across the network. However, physical constraints, hardware instabilities, and the inherently probabilistic nature of quantum sources often introduce fluctuations in generation timing. These variations can significantly affect synchronization between the source and the receiving nodes, potentially causing entangled photons to arrive outside the valid coincidence window and leading to missed detections. In monitoring systems, production jitter can be quantified by measuring the time differences between successive successful generation events and analyzing the variance or standard deviation of these intervals. Tracking production jitter over time is crucial for tuning the coincidence window and ensuring it remains wide enough to capture genuine events while minimizing accidental coincidences. In physical systems, production jitter can be measured by recording the time intervals between successive entanglement generation events. In simulations, it can be modeled by introducing timing uncertainty into the emission process, using Gaussian or Poisson distributions to represent hardware-induced variability.

\subsubsection{Coincidence Jitter}
\begin{figure}
    \centering
    \includegraphics[width=0.9\linewidth]{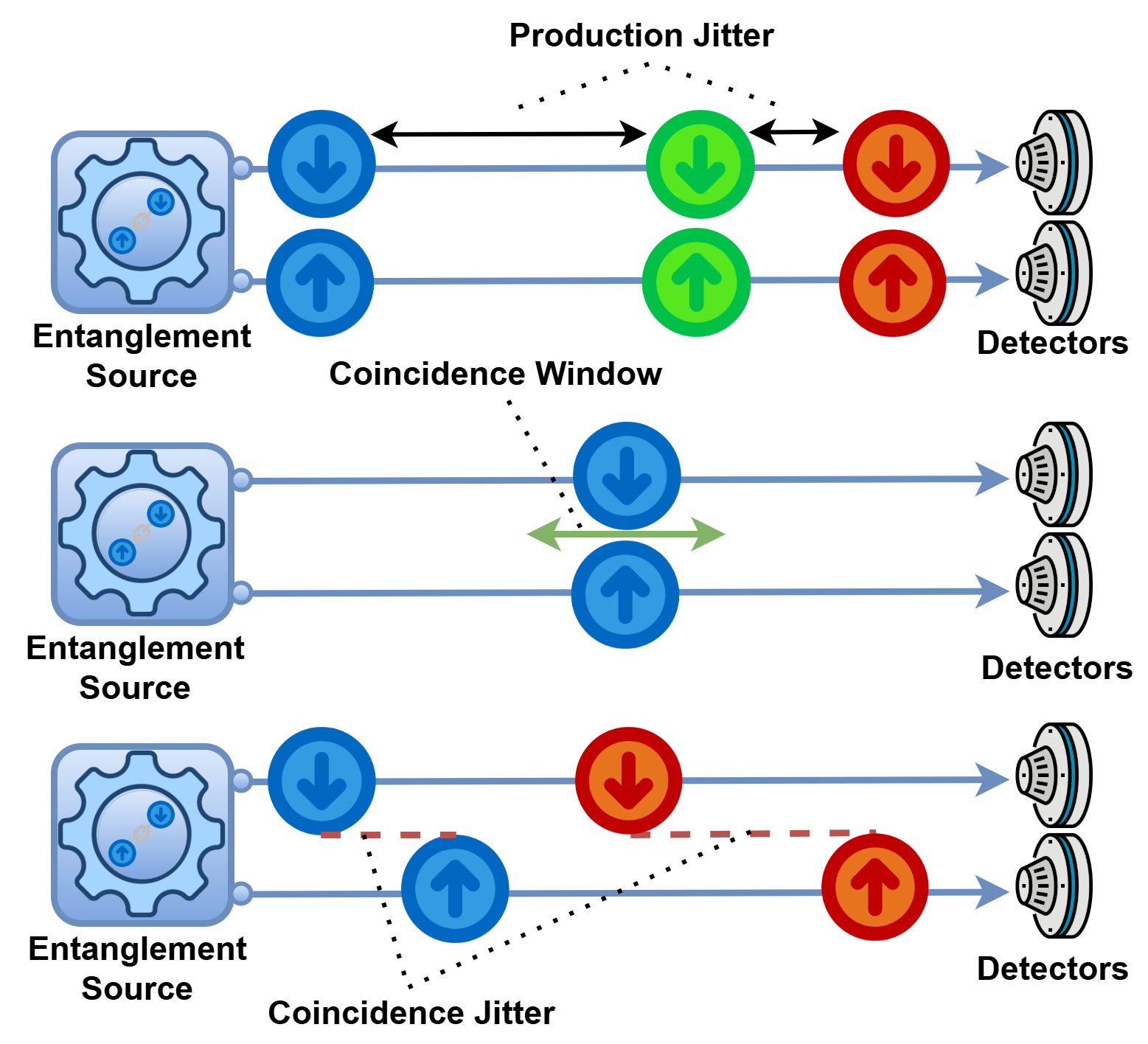}
    \caption{Illustration of timing metrics in quantum networks, showing an entanglement source emitting multiple photon pairs, each pair encoded with a distinct color to differentiate them, as they travel toward two detectors. This visualization highlights the differences between production jitter, coincidence jitter, and the coincidence window. Production jitter represents the variability in emission times between consecutive photon pairs. The coincidence window is the allowable time interval within which detection events are considered to originate from the same entangled pair. Coincidence jitter refers to the variation in arrival time differences between photons of the same pair at separate detectors.}
    \label{fig:timing-metrics}
\end{figure}

It refers to the variation in the time difference between two photons from the same entangled pair arriving at their respective detectors. Ideally, entangled photons are emitted simultaneously and travel along well-calibrated optical paths to reach their destinations within a tightly synchronized time window. Such calibration involves ensuring that the optical path lengths are equal, or that any differences are known and compensated for, and aligning components such as mirrors and lenses to minimize unintended delays. However, in real-world quantum networks, perfect synchronization is rarely achievable. Differences in channel length, chromatic dispersion in optical fibers (where different wavelengths propagate at different speeds), detector timing responses, and environmental disturbances such as temperature fluctuations or vibrations can all introduce timing variations between the photon arrivals. If the coincidence jitter exceeds the size of the configured coincidence window, the detection events may fall outside the acceptable time frame, causing valid entangled pairs to be disregarded. This results in missed detections, reduced entanglement rates, and overall degradation of network performance. Monitoring coincidence jitter in real time enables the dynamic adjustment of the coincidence window, allowing the system to adapt to current network conditions and mitigate performance loss. Such adaptability is vital in realistic deployments where environmental and network parameters are not fixed and can change unpredictably. In physical systems, coincidence jitter can be measured by analyzing the variation in arrival time differences between photons of the same entangled pair, using synchronized clocks and high-resolution timestamping hardware. In simulations, it can be modeled by introducing random arrival delays at the detectors, typically based on known hardware timing uncertainties or channel-specific propagation effects. The difference between the coincidence window, production jitter, and coincidence jitter is illustrated in Figure \ref{fig:timing-metrics}. 

\subsection{Exogenous Factors}
In addition to intrinsic quantum performance metrics, quantum networks require continuous monitoring of external environmental factors that significantly influence network performance, even though these quantities are not performance metrics themselves. In experimental platforms, exogenous factors can be measured using off-the-shelf sensors (e.g., thermometers, hygrometers, accelerometers) colocated with quantum components. In simulations, these factors are often not modeled directly, but their effects are incorporated through parameter adjustments (e.g., modifying loss, jitter, or dark count rate) to reflect realistic operating conditions.

\subsubsection{Source Temperature}
It refers to the operating temperature of the entangled photon source. Maintaining a stable source temperature is critical for ensuring consistent and high-quality entangled photon generation. Temperature fluctuations can alter the refractive index of the nonlinear medium, resulting in a shift in the phase-matching condition. This shift affects both the wavelength and the emission direction of the generated photons, potentially degrading the spectral and spatial overlap between entangled pairs and reducing entanglement fidelity. In addition, temperature instability can contribute to both production jitter, by affecting the timing consistency of photon generation, and coincidence jitter, by introducing variability in photon arrival times at the detectors. Monitoring the source temperature enables the detection of thermal drift and helps maintain the stability and reliability of the entangled photon source. 

\subsubsection{Room Temperature}
It refers to the ambient environmental temperature under which quantum devices operate. Fluctuations in room temperature can increase the dark count rate, affect mechanical alignment due to thermal expansion or contraction, and reduce optical path stability, particularly in FSO channels. Even minor thermal variations can expand optical mounts or fiber connectors, causing misalignment and introducing coincidence jitter. Temperature fluctuations can also alter the refractive index of optical channels, leading to drift in polarization and timing stability, which in turn affects entanglement fidelity and the dark count rate of detectors. Monitoring room temperature is therefore essential to maintain stable operating conditions and to support timely thermal compensation or realignment.

\subsubsection{Moisture}
Moisture, or atmospheric humidity, primarily affects FSO quantum communication. Water vapor in the air increases both scattering and absorption, leading to higher photon loss and a potential increase in decoherence due to interactions with environmental particles. High humidity can also damage sensitive optical components or coatings over time, especially in outdoor or semi-enclosed deployments. Increased humidity generally reduces transmission efficiency and degrades the visibility of interference patterns, impacting overall entanglement quality. Monitoring humidity is essential for preserving channel quality, minimizing optical degradation, and ensuring consistent quantum network performance.

\subsubsection{Vibrations}
It refers to mechanical disturbances that affect the physical stability of optical setups. These disturbances can cause mechanical misalignments, beam displacement in free-space systems, and connector instability in fiber-based links. Such effects may degrade signal quality, reduce entanglement fidelity, and increase the likelihood of timing errors. Monitoring vibrations helps determine whether adaptive optics or mechanical stabilization systems should be activated or tuned.

\vspace{0.2mm}
In addition to the core metrics discussed above, quantum testbeds often rely on an extended set of experimental metrics that support more detailed physical-layer characterization. These include parameters such as the coincidence-to-accidental ratio (CAR) for assessing signal-to-noise quality, polarization or Franson visibility for quantifying interference and phase stability, and CHSH or witness bounds for verifying non-classical correlations. 
Other useful observables include heralding efficiency, which reflects the quality of source-detector coupling, and Hong-Ou-Mandel (HOM) dip visibility, which measures photon indistinguishability, coherence time, which indicates the temporal stability of quantum states, and coupling efficiency, which characterizes photon transfer effectiveness between optical components or nodes. Unlike the operational metrics that require continuous monitoring, these extended observables are typically evaluated periodically, as they depend mainly on the alignment and health of the optical equipment. While not essential for routine operation, they provide valuable diagnostic insights into the physical stability and calibration of quantum network components. In addition, White Rabbit synchronization introduces metrics such as phase/time error, maximum time interval error (MTIE), time deviation (TDEV), and frequency drift. These synchronization indicators complement the extended set by quantifying sub-nanosecond clock alignment accuracy and providing context for timing-related variations observed across distributed quantum nodes.

While the proposed framework defines general categories and metrics, their expected ranges and relative importance depend strongly on both the underlying technology and the intended application. Current photonic quantum networks typically achieve entanglement fidelities between $0.85$ and $0.95$ \cite{Wehner2018} and QBER of approximately $1-5\%$, depending on channel distance and environmental stability. QKD prioritizes minimizing the QBER, since secure key generation requires values well below the theoretical threshold of about $11\%$ for BB84-based protocols \cite{Scarani_2009}. In contrast, distributed quantum computing and teleportation demand higher fidelities, often exceeding $0.99$, to preserve coherence across remote processors.  Environmental stability and synchronization precision are particularly critical for quantum sensing and metrology, where phase coherence and timing accuracy directly determine measurement precision \cite{K_m_r_2014}. Thus, while all metrics contribute to overall network performance, their prioritization varies across use cases, underscoring the importance of flexible observability tailored to both functionality and technology maturity.

\section{Use Cases and Applications}
\label{sec:applications}
The standardized metrics defined in this work offer a foundation for designing and implementing an integrated monitoring system for quantum networks. This monitoring system would continuously track the performance metrics to facilitate proactive maintenance and optimization. Furthermore, it can serve as the foundation for the development of a quantum network control system that dynamically adjusts network behavior based on real-time measurements.

\subsection{Real-Time Fault Diagnosis}
A monitoring system built on the proposed metrics can provide a comprehensive, real-time view of the quantum network's state. By tracking indicators such as entanglement fidelity, QBER, entanglement rate, and environmental conditions, the system can diagnose performance degradation, trigger alerts, and log historical data for analysis and benchmarking. This level of visibility is essential for maintaining network reliability and facilitating fault isolation in complex quantum infrastructures.

\subsection{Adaptive Timing Control}
Timing metrics such as production jitter and coincidence jitter can support adaptive control of detection parameters. For example, a control system can use real-time jitter measurements to dynamically adjust the coincidence window to maximize valid detections while minimizing false coincidences. Similarly, monitoring waiting time allows the network to detect service delays. Consequently, if the waiting time exceeds expected thresholds, the system can reprioritize entanglement requests to ensure fairness, meet QoS requirements, or adjust scheduling strategies to improve responsiveness and resource utilization.

\subsection{Routing and Resource Allocation}
In larger quantum networks, the proposed metrics can guide the selection of optimal entanglement paths and the allocation of quantum resources based on current network conditions. Metrics such as fidelity, entanglement rate, and waiting time each capture a different aspect of link performance and therefore influence routing in complementary ways. For instance, selecting paths based on entanglement rate favors high-throughput links but may overlook signal quality, whereas prioritizing fidelity ensures high-quality entanglement at the potential cost of reduced rate or increased latency. Similarly, waiting time captures the responsiveness of the protocol and indicates how quickly a service request can be fulfilled under current network conditions. These differences suggest that the metrics are complementary rather than redundant, with each highlighting a different aspect of network performance. Therefore, a robust routing strategy should consider multiple metrics jointly or adaptively prioritize them based on the application's requirements, enabling more reliable and efficient use of quantum resources.

\subsection{Benchmarking and Comparative Evaluation}
The defined metrics also serve as standardized performance indicators for both simulated and physical quantum networks. Researchers and system designers can use these metrics for benchmarking purposes, such as comparing the performance of different protocols, hardware configurations, or network architectures. The adoption of a common metric set promotes consistency and comparability across platforms, simulations, and experimental setups, thereby accelerating progress in quantum network design and evaluation.

\subsection{Autonomous Quantum Network Control}
The monitoring system can serve as the foundation for developing a quantum network control plane capable of autonomously responding to environmental and performance conditions. For example, temperature changes could automatically trigger recalibration routines for the entangled photon source, while variations in ambient humidity could prompt adjustments to optical alignment systems to compensate for environmentally induced signal degradation. This type of closed-loop control enables dynamic system tuning without human intervention. Such a control plane also supports the development of quantum software-defined networks (QSDNs), where operational parameters are continually refined based on real-time feedback from standardized network metrics.

\section{Real-World Monitoring Prototype at ORNL}
\label{sec:prototype}
To explore the feasibility of building a comprehensive quantum network monitoring and control system, we developed an initial prototype focused on non-invasive monitoring of exogenous factors, specifically temperature and humidity, which can significantly affect quantum network stability and performance. This prototype serves as a foundational step toward our broader goal of creating a complete real-time observability and adaptive control framework for quantum networks. The monitoring system employs a Raspberry Pi 5 connected to an SHT35 digital sensor, with measurements collected every $5$ seconds to monitor ambient temperature and humidity conditions around the experimental setup. In addition, the temperature of the entangled photon source is monitored through the laser diode mount to observe variations in source operating conditions. The collected data is pushed to a Prometheus time-series database and visualized through a Grafana dashboard. To ensure experimental reliability and support future adaptive control, alert thresholds are defined such that temperature alerts are triggered if the ambient temperature exceeds \SI{23}{\celsius}, while humidity alerts are raised if the relative humidity exceeds $60\%$ or drops below $20\%$. Moreover, if the sensor fails or disconnects, the system automatically raises an alert in the Grafana dashboard, signaling a data acquisition issue. Figure \ref{fig:network} shows the prototype setup and a sample visualization of the monitoring dashboard.

\subsection{Impact of Source Temperature on Network Performance}
\begin{figure}
    \centering
    \includegraphics[width=1\linewidth]{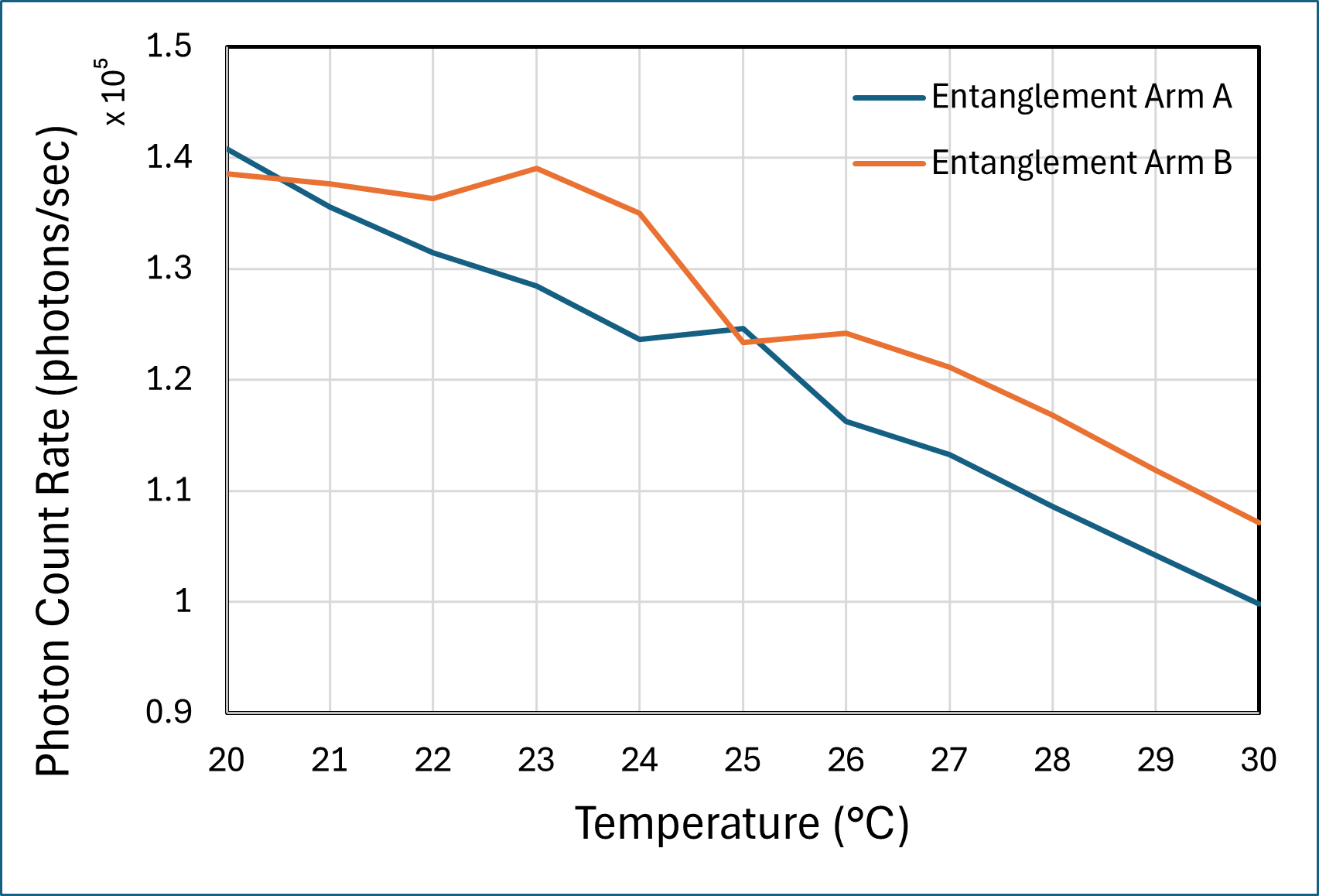}
    \caption{Effect of source temperature on photon detection rates in the two entanglement arms.} 
    \label{fig:count_rate}
\end{figure}

\begin{figure}
    \centering
    \includegraphics[width=1\linewidth]{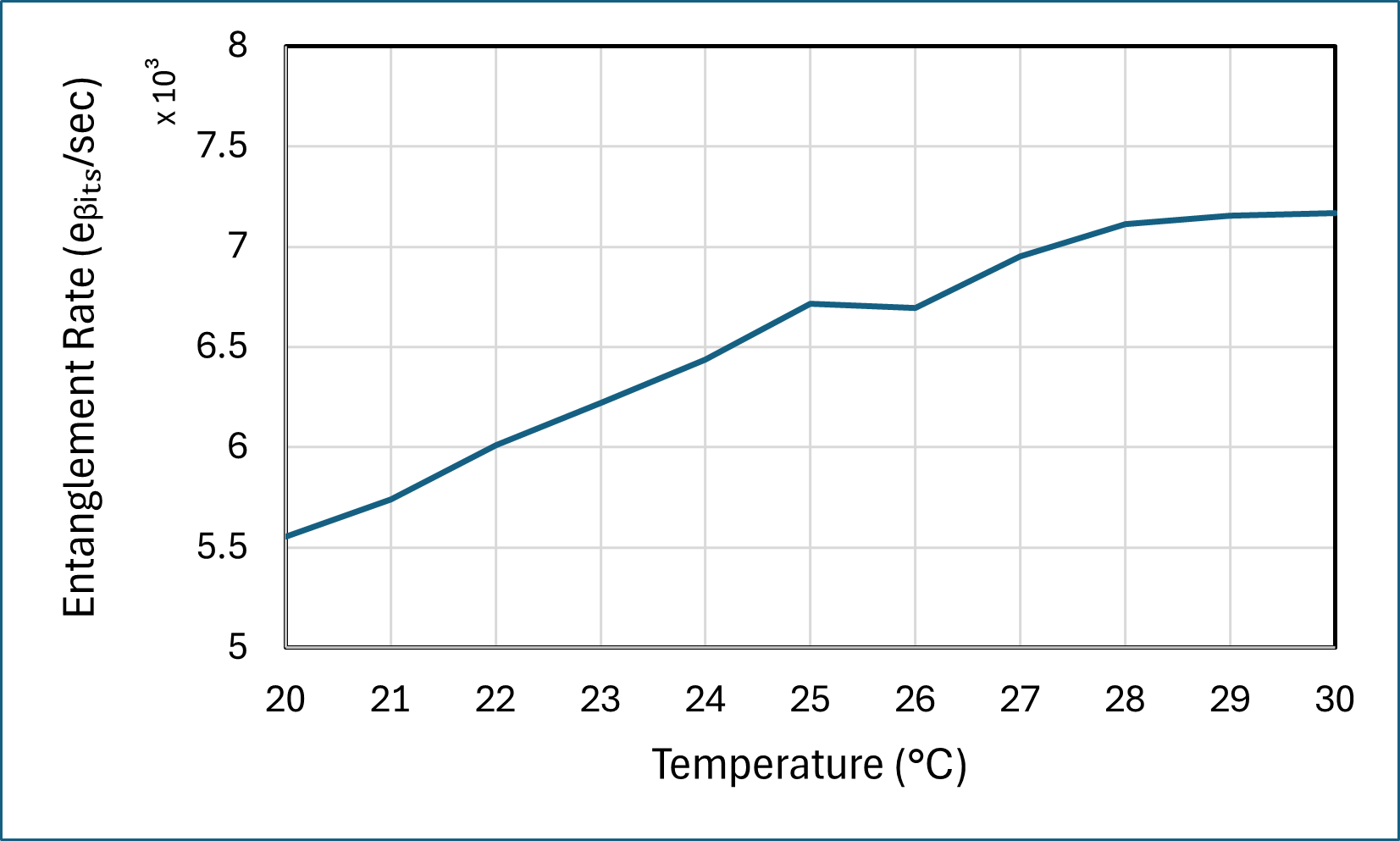}
    \caption{Effect of source temperature on entanglement rate measured through coincidence events between the two entanglement arms.}
    \label{fig:ent_rate}
\end{figure}

To illustrate the effect of exogenous factors on quantum network performance, we conducted an experiment analyzing the impact of source temperature on photon detection rates and entanglement rate. The source temperature was varied between \SI{20}{\celsius} and \SI{30}{\celsius} while recording photon count rates from the two entanglement arms as well as coincidence events between them. Each measurement point represents the average count and coincidence rates over a 100-second observation window. As shown in Figure \ref{fig:count_rate}, the photon detection rates in the two arms decreased as the source temperature increased, dropping from approximately $1.41 \times 10^5$ to $9.98 \times 10^4$ photons/sec in one arm and from $1.39 \times 10^5$ to $1.07 \times 10^5$ photons/sec in the other, corresponding to reductions of roughly $29\%$ and $23\%$, respectively. In contrast, the entanglement rate increased from about $5.5 \times 10^3$ to $7.2 \times 10^3$ $e_{bits}/sec$, as illustrated in Figure \ref{fig:ent_rate}. These observations demonstrate that variations in source temperature can significantly influence photon detection rates and entanglement generation behavior. Even small temperature fluctuations can produce measurable effects. For example, a change of only \SI{0.1}{\celsius} around \SI{23}{\celsius} resulted in approximately a $1.5 \times 10^3$ reduction in photon count rate and an increase of about $40$ to $50$ entanglement events in our setup. These results highlight the importance of monitoring both exogenous factors and quantum network performance metrics to better understand and manage variations in operational quantum networks.

\section{Challenges}
\label{sec:challenges}
The proposed metrics provide a structured and standardized foundation for monitoring and controlling quantum networks, but their effective adoption faces several practical challenges that must be carefully addressed.

\subsection{Quantum-Layer Monitoring Limitations}
Monitoring quantum networks presents several unique technical challenges that do not arise in classical systems. First, quantum measurements are inherently destructive, preventing direct inspection of transmitted quantum states. Estimating quantities such as entanglement fidelity or QBER requires sampling or reconstructing the state indirectly through repeated measurements. Full tomography is particularly costly, as it may require dozens of measurement settings and consumes significant photon resources. Second, monitoring often reduces the available quantum signal. For example, splitting photon streams to observe timing or spectral properties decreases the probability of successful entanglement generation and lowers end-to-end rates. Third, accurate timing characterization requires sub-nanosecond synchronization and calibrated delay compensation across heterogeneous channels. Small drifts in fiber length, detector response, or environmental conditions can distort timing histograms and complicate coincidence analysis. Additionally, quantum links are highly sensitive to environmental variations such as temperature, humidity, and vibration, yet no standard telemetry framework exists for correlating these external factors with quantum performance. This creates a critical trade-off between observability and performance. While detailed monitoring can yield richer insights into system behavior, it may also reduce the availability or quality of quantum resources. Practical deployments must carefully balance observability with system performance.

\subsection{Operational Overhead of Monitoring}
Even when appropriate measurement techniques exist, incorporating continuous monitoring into a running quantum network introduces operational overhead that must be carefully managed. Collecting performance metrics requires additional data acquisition, timestamping, buffering, and communication across nodes, all of which consume computational resources on the control hardware.  Monitoring processes may trigger calibration routines, metadata exchanges, or synchronization updates that momentarily interrupt or slow normal network operation. Similarly, generating, aggregating, and transporting monitoring data increases classical communication load, which may introduce latency in software-defined control loops that depend on timely feedback.

\subsection{Metric Interdependencies}
Many metrics are not independent. For instance, a rise in temperature may simultaneously increase dark count rate, introduce path-length fluctuations, and contribute to timing jitter, all of which can affect entanglement fidelity and rate. Similarly, loss, QBER, and visibility are often correlated, since physical disturbances that affect one metric typically influence several others simultaneously. This interdependence complicates diagnosis and demands a monitoring system capable of correlating multiple metrics to isolate root causes of performance degradation.

\subsection{Control Complexity and System Constraints}
Although this work focuses on observability, the defined metrics can support autonomous control systems that adaptively tune network parameters in response to real-time conditions. Examples include tuning pump power, adjusting polarization controllers, modifying coincidence windows, or switching entanglement routes. However, building such closed-loop control mechanisms requires reliable measurements, stability guarantees, and the ability to deal with uncertainty and probabilistic behavior, challenges not yet fully addressed in the current stage of quantum network development.

\section{Conclusion and Future Work}
\label{sec:conclusion}
This article proposed a structured framework for monitoring quantum networks through a set of standardized performance metrics. The framework supports real-time observability, benchmarking, and lays the foundation for quantum network control systems and quantum software-defined networking. To demonstrate the feasibility of the framework, we implemented a prototype environmental monitoring system integrated with the quantum network infrastructure at ORNL. The system monitors ambient temperature and humidity using a Raspberry Pi 5, SHT35 sensor, Prometheus database, and Grafana dashboard, and provides real-time alerts to ensure experimental reliability. This initial deployment validates the practicality of integrating monitoring capabilities into operational quantum networks and serves as a foundational step toward a comprehensive monitoring and control system. Although the proposed framework offers many advantages, its practical implementation presents several challenges. For example, measuring certain metrics in real time, such as fidelity and QBER, may interfere with active quantum protocols and require trade-offs between observability and resource constraints. Additionally, metric interdependencies and the probabilistic nature of quantum systems present open challenges for robust control design. To advance this vision, future work will focus on the following directions.

\subsection{Extending the Monitoring Prototype}
Extending prototype monitoring systems to include a broader set of quantum performance metrics, such as quality metrics, throughput and latency metrics, and timing metrics, remains an important research direction. Doing so requires standardized interfaces for accessing quantum-layer information from photon sources, detectors, and timing hardware, as well as methods for synchronizing and aggregating this data across distributed nodes. Advancing these architectural and integration capabilities will enable future monitoring systems to provide unified visibility across both the classical and quantum layers of a network, forming a foundation for real-time diagnostics, performance optimization, and adaptive control.

\subsection{Low-Overhead Metric Collection}
Certain quantum metrics, especially those requiring direct measurement of quantum states, can disrupt ongoing quantum operations and degrade network performance. To address this, we will investigate non-intrusive, low-overhead monitoring techniques that rely on passive observation or indirect inference. These approaches aim to maintain high observability while minimizing interference.

\subsection{Correlation and Minimal Observability}
Future work will focus on analyzing correlations among performance metrics and developing techniques to identify and model these relationships. Understanding how key metrics such as fidelity, QBER, loss, and timing jitter influence each other will improve the interpretability of network performance and enhance decision-making for diagnostics and control. Another open direction is investigating whether a smaller subset of the proposed metrics can provide sufficient visibility into network health and performance. Many of the defined metrics are complementary rather than redundant. For example, entanglement fidelity and QBER both reflect link integrity, but fidelity captures quantum state accuracy, while QBER reflects bit-level correctness after measurement. Similarly, exogenous factors such as temperature may affect dark count rate, which in turn impacts fidelity and entanglement rate. These interdependencies suggest that collecting all metrics may be unnecessary in some contexts. Future efforts should aim to formally model these relationships and identify a core set of metrics that achieve strong observability with minimal overhead.

\subsection{Predictive Network Monitoring}
Future work should also focus on integrating machine learning models capable of predicting performance degradation and recommending corrective actions. By analyzing time-series data collected from key performance metrics, these models can anticipate failures, suggest optimal configuration changes, or trigger dynamic rerouting decisions. Such predictive monitoring is essential for enabling autonomous control and improving the resilience and efficiency of quantum networks.

\section{ACKNOWLEDGMENTS}
This manuscript has been authored by UT-Battelle, LLC, under contract DE-AC05-00OR22725 with the US Department of Energy (DOE). The US government retains and the publisher, by accepting the article for publication, acknowledges that the US government retains a nonexclusive, paid-up, irrevocable, worldwide license to publish or reproduce the published form of this manuscript or allow others to do so, for US government purposes. DOE will provide public access to these results of federally sponsored research in accordance with the DOE Public Access Plan (\url{https://www.energy.gov/downloads/doe-public-access-plan}).

This work was supported by the U.S. DOE Office of Science, Office of Advanced Scientific Computing Research Early Career Grant ``Large Scale Deep Learning for Intelligent Networks'' award ERKJ435 hosted at Oak Ridge National Laboratory.

\printcredits

\bibliographystyle{cas-model2-names}

\bibliography{References}

\begin{thebibliography}{23}
\expandafter\ifx\csname natexlab\endcsname\relax\def\natexlab#1{#1}\fi
\providecommand{\url}[1]{\texttt{#1}}
\providecommand{\href}[2]{#2}
\providecommand{\path}[1]{#1}
\providecommand{\DOIprefix}{doi:}
\providecommand{\ArXivprefix}{arXiv:}
\providecommand{\URLprefix}{URL: }
\providecommand{\Pubmedprefix}{pmid:}
\providecommand{\doi}[1]{\href{http://dx.doi.org/#1}{\path{#1}}}
\providecommand{\Pubmed}[1]{\href{pmid:#1}{\path{#1}}}
\providecommand{\bibinfo}[2]{#2}
\ifx\xfnm\relax \def\xfnm[#1]{\unskip,\space#1}\fi
\bibitem[{Abane et~al.(2025)Abane, Cubeddu, Mai and Battou}]{10882978}
\bibinfo{author}{Abane, A.}, \bibinfo{author}{Cubeddu, M.}, \bibinfo{author}{Mai, V.S.}, \bibinfo{author}{Battou, A.}, \bibinfo{year}{2025}.
\newblock \bibinfo{title}{Entanglement routing in quantum networks: A comprehensive survey}.
\newblock \bibinfo{journal}{IEEE Transactions on Quantum Engineering} \bibinfo{volume}{6}, \bibinfo{pages}{1--39}.
\newblock \DOIprefix\doi{10.1109/TQE.2025.3541123}.
\bibitem[{Alshowkan et~al.(2021)Alshowkan, Williams, Evans, Rao, Simmerman, Lu, Lingaraju, Weiner, Marvinney, Pai, Lawrie, Peters and Lukens}]{PRXQuantum.2.040304}
\bibinfo{author}{Alshowkan, M.}, \bibinfo{author}{Williams, B.P.}, \bibinfo{author}{Evans, P.G.}, \bibinfo{author}{Rao, N.S.}, \bibinfo{author}{Simmerman, E.M.}, \bibinfo{author}{Lu, H.H.}, \bibinfo{author}{Lingaraju, N.B.}, \bibinfo{author}{Weiner, A.M.}, \bibinfo{author}{Marvinney, C.E.}, \bibinfo{author}{Pai, Y.Y.}, \bibinfo{author}{Lawrie, B.J.}, \bibinfo{author}{Peters, N.A.}, \bibinfo{author}{Lukens, J.M.}, \bibinfo{year}{2021}.
\newblock \bibinfo{title}{Reconfigurable quantum local area network over deployed fiber}.
\newblock \bibinfo{journal}{PRX Quantum} \bibinfo{volume}{2}, \bibinfo{pages}{040304}.
\newblock \URLprefix \url{https://link.aps.org/doi/10.1103/PRXQuantum.2.040304}, \DOIprefix\doi{10.1103/PRXQuantum.2.040304}.
\bibitem[{Bennett et~al.(1993)Bennett, Brassard, Cr\'epeau, Jozsa, Peres and Wootters}]{PhysRevLett.70.1895}
\bibinfo{author}{Bennett, C.H.}, \bibinfo{author}{Brassard, G.}, \bibinfo{author}{Cr\'epeau, C.}, \bibinfo{author}{Jozsa, R.}, \bibinfo{author}{Peres, A.}, \bibinfo{author}{Wootters, W.K.}, \bibinfo{year}{1993}.
\newblock \bibinfo{title}{Teleporting an unknown quantum state via dual classical and einstein-podolsky-rosen channels}.
\newblock \bibinfo{journal}{Phys. Rev. Lett.} \bibinfo{volume}{70}, \bibinfo{pages}{1895--1899}.
\newblock \URLprefix \url{https://link.aps.org/doi/10.1103/PhysRevLett.70.1895}, \DOIprefix\doi{10.1103/PhysRevLett.70.1895}.
\bibitem[{Beukers et~al.(2024)Beukers, Pasini, Choi, Englund, Hanson and Borregaard}]{PRXQuantum.5.010202}
\bibinfo{author}{Beukers, H.K.}, \bibinfo{author}{Pasini, M.}, \bibinfo{author}{Choi, H.}, \bibinfo{author}{Englund, D.}, \bibinfo{author}{Hanson, R.}, \bibinfo{author}{Borregaard, J.}, \bibinfo{year}{2024}.
\newblock \bibinfo{title}{Remote-entanglement protocols for stationary qubits with photonic interfaces}.
\newblock \bibinfo{journal}{PRX Quantum} \bibinfo{volume}{5}, \bibinfo{pages}{010202}.
\newblock \URLprefix \url{https://link.aps.org/doi/10.1103/PRXQuantum.5.010202}, \DOIprefix\doi{10.1103/PRXQuantum.5.010202}.
\bibitem[{Cacciapuoti et~al.(2020)Cacciapuoti, Caleffi, Tafuri, Cataliotti, Gherardini and Bianchi}]{8910635}
\bibinfo{author}{Cacciapuoti, A.S.}, \bibinfo{author}{Caleffi, M.}, \bibinfo{author}{Tafuri, F.}, \bibinfo{author}{Cataliotti, F.S.}, \bibinfo{author}{Gherardini, S.}, \bibinfo{author}{Bianchi, G.}, \bibinfo{year}{2020}.
\newblock \bibinfo{title}{Quantum internet: Networking challenges in distributed quantum computing}.
\newblock \bibinfo{journal}{IEEE Network} \bibinfo{volume}{34}, \bibinfo{pages}{137--143}.
\newblock \DOIprefix\doi{10.1109/MNET.001.1900092}.
\bibitem[{Caleffi(2017)}]{8068178}
\bibinfo{author}{Caleffi, M.}, \bibinfo{year}{2017}.
\newblock \bibinfo{title}{Optimal routing for quantum networks}.
\newblock \bibinfo{journal}{IEEE Access} \bibinfo{volume}{5}, \bibinfo{pages}{22299--22312}.
\newblock \DOIprefix\doi{10.1109/ACCESS.2017.2763325}.
\bibitem[{Cerutti et~al.(2023)Cerutti, Lewis and Bonavitacola}]{JRC132426}
\bibinfo{author}{Cerutti, I.}, \bibinfo{author}{Lewis, A.}, \bibinfo{author}{Bonavitacola, F.}, \bibinfo{year}{2023}.
\newblock \bibinfo{title}{Quantum Key Distribution ({QKD}): Experimental Assessment}.
\newblock \bibinfo{type}{Technical Report} \bibinfo{number}{JRC132426}. Joint Research Centre (JRC), European Commission. \bibinfo{address}{Luxembourg}.
\newblock \DOIprefix\doi{10.2760/804200}.
\bibitem[{Choi et~al.(2004)Choi, Moon, Zhang, Papagiannaki and Diot}]{1354590}
\bibinfo{author}{Choi, B.K.}, \bibinfo{author}{Moon, S.}, \bibinfo{author}{Zhang, Z.L.}, \bibinfo{author}{Papagiannaki, K.}, \bibinfo{author}{Diot, C.}, \bibinfo{year}{2004}.
\newblock \bibinfo{title}{Analysis of point-to-point packet delay in an operational network}, in: \bibinfo{booktitle}{IEEE INFOCOM 2004}, pp. \bibinfo{pages}{1797--1807 vol.3}.
\newblock \DOIprefix\doi{10.1109/INFCOM.2004.1354590}.
\bibitem[{Dahlberg et~al.(2019)Dahlberg, Skrzypczyk, Coopmans, Wubben, Rozpundefineddek, Pompili, Stolk, Pawe\l{}czak, Knegjens, de~Oliveira~Filho, Hanson and Wehner}]{10.1145/3341302.3342070}
\bibinfo{author}{Dahlberg, A.}, \bibinfo{author}{Skrzypczyk, M.}, \bibinfo{author}{Coopmans, T.}, \bibinfo{author}{Wubben, L.}, \bibinfo{author}{Rozpundefineddek, F.}, \bibinfo{author}{Pompili, M.}, \bibinfo{author}{Stolk, A.}, \bibinfo{author}{Pawe\l{}czak, P.}, \bibinfo{author}{Knegjens, R.}, \bibinfo{author}{de~Oliveira~Filho, J.}, \bibinfo{author}{Hanson, R.}, \bibinfo{author}{Wehner, S.}, \bibinfo{year}{2019}.
\newblock \bibinfo{title}{A link layer protocol for quantum networks}, in: \bibinfo{booktitle}{Proceedings of the ACM Special Interest Group on Data Communication}, \bibinfo{publisher}{Association for Computing Machinery}, \bibinfo{address}{New York, NY, USA}. p. \bibinfo{pages}{159–173}.
\newblock \URLprefix \url{https://doi.org/10.1145/3341302.3342070}, \DOIprefix\doi{10.1145/3341302.3342070}.
\bibitem[{{ESnet}(2025)}]{iperf}
\bibinfo{author}{{ESnet}}, \bibinfo{year}{2025}.
\newblock \bibinfo{title}{{iPerf3: A TCP, UDP, and SCTP Network Bandwidth Measurement Tool}}.
\newblock \bibinfo{howpublished}{\url{https://software.es.net/iperf/}}.
\newblock \bibinfo{note}{Accessed: June 4, 2025}.
\bibitem[{Fitzsimons(2017)}]{Fitzsimons2017}
\bibinfo{author}{Fitzsimons, J.F.}, \bibinfo{year}{2017}.
\newblock \bibinfo{title}{Private quantum computation: an introduction to blind quantum computing and related protocols}.
\newblock \bibinfo{journal}{npj Quantum Information} \bibinfo{volume}{3}, \bibinfo{pages}{23}.
\newblock \URLprefix \url{https://doi.org/10.1038/s41534-017-0025-3}, \DOIprefix\doi{10.1038/s41534-017-0025-3}.
\bibitem[{Jozsa(1994)}]{136}
\bibinfo{author}{Jozsa, R.}, \bibinfo{year}{1994}.
\newblock \bibinfo{title}{Fidelity for mixed quantum states}.
\newblock \bibinfo{journal}{Journal of Modern Optics} \bibinfo{volume}{41}, \bibinfo{pages}{2315--2323}.
\newblock \URLprefix \url{https://doi.org/10.1080/09500349414552171}, \DOIprefix\doi{10.1080/09500349414552171}, \href{http://arxiv.org/abs/https://doi.org/10.1080/09500349414552171}{\tt arXiv:https://doi.org/10.1080/09500349414552171}.
\bibitem[{Kómár et~al.(2014)Kómár, Kessler, Bishof, Jiang, Sørensen, Ye and Lukin}]{K_m_r_2014}
\bibinfo{author}{Kómár, P.}, \bibinfo{author}{Kessler, E.M.}, \bibinfo{author}{Bishof, M.}, \bibinfo{author}{Jiang, L.}, \bibinfo{author}{Sørensen, A.S.}, \bibinfo{author}{Ye, J.}, \bibinfo{author}{Lukin, M.D.}, \bibinfo{year}{2014}.
\newblock \bibinfo{title}{A quantum network of clocks}.
\newblock \bibinfo{journal}{Nature Physics} \bibinfo{volume}{10}, \bibinfo{pages}{582–587}.
\newblock \URLprefix \url{http://dx.doi.org/10.1038/nphys3000}, \DOIprefix\doi{10.1038/nphys3000}.
\bibitem[{Mah()}]{nuttcp}
\bibinfo{author}{Mah, B.A.}, .
\newblock \bibinfo{title}{{nuttcp}: A network performance measurement tool}.
\newblock \bibinfo{howpublished}{\url{https://www.nuttcp.net/}}.
\newblock \bibinfo{note}{Accessed: June 4, 2025}.
\bibitem[{Menkart et~al.(2022)Menkart, Hart, Murphy and Roy}]{Menkart:22}
\bibinfo{author}{Menkart, N.}, \bibinfo{author}{Hart, J.D.}, \bibinfo{author}{Murphy, T.E.}, \bibinfo{author}{Roy, R.}, \bibinfo{year}{2022}.
\newblock \bibinfo{title}{Dark current and single photon detection by 1550 nm avalanche photodiodes: dead time corrected probability distributions and entropy rates}.
\newblock \bibinfo{journal}{Opt. Express} \bibinfo{volume}{30}, \bibinfo{pages}{39431--39444}.
\newblock \URLprefix \url{https://opg.optica.org/oe/abstract.cfm?URI=oe-30-22-39431}, \DOIprefix\doi{10.1364/OE.466330}.
\bibitem[{Scarani et~al.(2009)Scarani, Bechmann-Pasquinucci, Cerf, Dušek, Lütkenhaus and Peev}]{Scarani_2009}
\bibinfo{author}{Scarani, V.}, \bibinfo{author}{Bechmann-Pasquinucci, H.}, \bibinfo{author}{Cerf, N.J.}, \bibinfo{author}{Dušek, M.}, \bibinfo{author}{Lütkenhaus, N.}, \bibinfo{author}{Peev, M.}, \bibinfo{year}{2009}.
\newblock \bibinfo{title}{The security of practical quantum key distribution}.
\newblock \bibinfo{journal}{Reviews of Modern Physics} \bibinfo{volume}{81}, \bibinfo{pages}{1301–1350}.
\newblock \URLprefix \url{http://dx.doi.org/10.1103/RevModPhys.81.1301}, \DOIprefix\doi{10.1103/revmodphys.81.1301}.
\bibitem[{Shaban et~al.(2024a)Shaban, Ismail and Kiran}]{10820730}
\bibinfo{author}{Shaban, M.}, \bibinfo{author}{Ismail, M.}, \bibinfo{author}{Kiran, M.}, \bibinfo{year}{2024}a.
\newblock \bibinfo{title}{{QNTN}: Establishing a regional quantum network in tennessee}, in: \bibinfo{booktitle}{SC24-W: Workshops of the International Conference for High Performance Computing, Networking, Storage and Analysis}, pp. \bibinfo{pages}{810--818}.
\newblock \DOIprefix\doi{10.1109/SCW63240.2024.00115}.
\bibitem[{Shaban et~al.(2024b)Shaban, Ismail and Saad}]{10684482}
\bibinfo{author}{Shaban, M.}, \bibinfo{author}{Ismail, M.}, \bibinfo{author}{Saad, W.}, \bibinfo{year}{2024}b.
\newblock \bibinfo{title}{{SPARQ}: Efficient entanglement distribution and routing in space–air–ground quantum networks}.
\newblock \bibinfo{journal}{IEEE Transactions on Quantum Engineering} \bibinfo{volume}{5}, \bibinfo{pages}{1--20}.
\newblock \DOIprefix\doi{10.1109/TQE.2024.3464572}.
\bibitem[{Shettell et~al.(2022)Shettell, Hassani and Markham}]{shettell}
\bibinfo{author}{Shettell, N.}, \bibinfo{author}{Hassani, M.}, \bibinfo{author}{Markham, D.}, \bibinfo{year}{2022}.
\newblock \bibinfo{title}{Private network parameter estimation with quantum sensors}.
\newblock \URLprefix \url{https://arxiv.org/abs/2207.14450}, \href{http://arxiv.org/abs/2207.14450}{\tt arXiv:2207.14450}.
\bibitem[{Tierney et~al.(2009)Tierney, Metzger, Boote, Boyd, Brown, Carlson, Zekauskas, Zurawski, Swany and Grigoriev}]{tierney2009perfsonar}
\bibinfo{author}{Tierney, B.}, \bibinfo{author}{Metzger, J.}, \bibinfo{author}{Boote, J.}, \bibinfo{author}{Boyd, E.}, \bibinfo{author}{Brown, A.}, \bibinfo{author}{Carlson, R.}, \bibinfo{author}{Zekauskas, M.}, \bibinfo{author}{Zurawski, J.}, \bibinfo{author}{Swany, M.}, \bibinfo{author}{Grigoriev, M.}, \bibinfo{year}{2009}.
\newblock \bibinfo{title}{perfsonar: Instantiating a global network measurement framework}.
\newblock \bibinfo{journal}{SOSP Wksp. Real Overlays and Distrib. Sys} \bibinfo{volume}{28}.
\bibitem[{Vardoyan and Wehner(2023)}]{10313675}
\bibinfo{author}{Vardoyan, G.}, \bibinfo{author}{Wehner, S.}, \bibinfo{year}{2023}.
\newblock \bibinfo{title}{Quantum network utility maximization}, in: \bibinfo{booktitle}{2023 IEEE International Conference on Quantum Computing and Engineering (QCE)}, pp. \bibinfo{pages}{1238--1248}.
\newblock \DOIprefix\doi{10.1109/QCE57702.2023.00140}.
\bibitem[{Wehner et~al.(2018)Wehner, Elkouss and Hanson}]{Wehner2018}
\bibinfo{author}{Wehner, S.}, \bibinfo{author}{Elkouss, D.}, \bibinfo{author}{Hanson, R.}, \bibinfo{year}{2018}.
\newblock \bibinfo{title}{Quantum internet: A vision for the road ahead}.
\newblock \bibinfo{journal}{Science} \bibinfo{volume}{362}, \bibinfo{pages}{eaam9288}.
\newblock \DOIprefix\doi{10.1126/science.aam9288}.
\bibitem[{Wu et~al.(2023)Wu, Matsui, Forrer, Soeda, Andrés-Martínez, Mills, Henaut and Murao}]{146}
\bibinfo{author}{Wu, J.Y.}, \bibinfo{author}{Matsui, K.}, \bibinfo{author}{Forrer, T.}, \bibinfo{author}{Soeda, A.}, \bibinfo{author}{Andrés-Martínez, P.}, \bibinfo{author}{Mills, D.}, \bibinfo{author}{Henaut, L.}, \bibinfo{author}{Murao, M.}, \bibinfo{year}{2023}.
\newblock \bibinfo{title}{Entanglement-efficient bipartite-distributed quantum computing}.
\newblock \bibinfo{journal}{Quantum} \bibinfo{volume}{7}, \bibinfo{pages}{1196}.
\newblock \DOIprefix\doi{10.22331/q-2023-12-05-1196}.

\end{thebibliography}


\end{document}